\documentstyle[aps,prl,twocolumn,epsf]{revtex}
\begin{document}
\draft
\preprint{}
\twocolumn[\hsize\textwidth\columnwidth\hsize\csname @twocolumnfalse\endcsname

\title{Wavelength Teleportation via Distant Quantum Entanglement}
\author{M.S. Shahriar$^{1,2}$, P. Pradhan$^{1}$, V. Gopal$^{1}$, G.S. Pati$^{1}$, and  G.C. Cardoso$^{1}$ }
\address{$^{1}$Department  of Electrical and Computer Engineering, Northwestern Univeristy,
    Evanston, IL 60208  \\
   $^{2}$Research Laboratory of Electronics, Massachusetts Institute of Technology,
 Cambridge, MA 02139 }
\maketitle
\date{\today}
\begin{abstract}
Recently, we have shown theoretically \cite{Shahriar02} as well
as experimentally \cite{Cardoso03} how the 
phase of an electromagnetic field can be determined by measuring the 
population of either of the two states of a two-level atomic system 
excited by this field, via the so-called Bloch-Siegert oscillation 
resulting from the interference between the co- and counter-rotating 
excitations.  Here, we show how a degenerate entanglement, created without 
transmitting any timing signal, can be used to teleport this phase 
information.  This phase-teleportation process may be applied to achieve 
wavelength teleportation, which in turn may be used for frequency-locking 
of remote oscillators.   
\end{abstract} 
\pacs{ 03.67.-a, 03.67.Hk, 03.67.Lx, 32.80.Qk } 
]
The task of synchronizing a pair of clocks that are separated in space is important for many 
practical applications, such as  global positioning systems (GPS) \cite{Misra01}    and  very large base 
interferometry (VLBI) \cite{Lavy87}.  Conventionally, the synchronization is performed by transmitting 
timing signals between the clocks.  Consider first the ideal situation where the intervening medium is 
stable and fully characterized.  
The accuracy of the synchronization process is  then limited by the uncertainty in the timing 
signal.  The best result achievable is limited by the 
signal to noise ratio (SNR). Given enough resources, one can eliminate sources of systematic 
noises, so that the fundamental constraint is the shot noise limit (SNL).
 In principle, specially prepared quantum states can reduce 
the effective noise below the SNL.  
However, 
the actual SNR achievable this way is far 
below what can be achieved using classical states.   
Most of the recent proposals \cite{Chuang00,Giovannetti01,Giovannetti01a,Giovannetti02,Jozsa00}
for achieving improved oscillator synchronization (OS) using quantum processes suffer from the same constraint, 
so that in  practice they are inferior to classical approaches.  Thus, given the current state of 
technology, quantum mechanical effects is not likely to help in the process of OS  under the 
{\em ideal situation}.  

  Consider next the {\em realistic} situation where the density of the intervening medium 
fluctuates randomly, leading to a corresponding fluctuation in the time 
needed for a signal to travel between the clocks.  Under this condition, it is 
{\em fundamentally impossible}, barring superluminal  propagation, to synchronize the clocks 
exactly.  This follows from the principle of special  relativity, which is built on the axiom that 
there exists a maximum speed --- namely, the speed of  light in vacuum --- at which information 
can propagate.  As such, the notion of clock  synchrony is  {\em defined} with respect to the time 
it takes for light to traverse the distance between the  clocks. It then follows that if this 
travel time itself is fluctuating, then the clock synchrony is 
undefined, and cannot be achieved on the timescale of the fluctuation. One can define and establish 
only an average synchrony, valid only for timescales longer than that of the fluctuation. 
In all situations of practical interest, OS  always implies the achievement of this 
{\em average synchrony}.  

  An alternative way to improve the average synchrony is through frequency locking.  
 Specifically, consider a typical application where each oscillator  
is locked to a metastable atomic transition.  Most of the 
recent  proposals about clock synchronization, including the Jozsa protocol \cite{Jozsa00}, make the 
assumption  that each oscillator continues to operate at some ideal transition frequency.  
In practice, however, this is not the case.  The  frequency of each oscillator 
undergoes shifts and drifts due to a host of reasons.  These fluctuations lie at the 
heart of clock asynchrony.  As such, minimizing the relative drifts in the 
frequencies is perhaps the most effective way to minimize the error in OS. 
This approach opens up new possibilities for exploring whether quantum  mechanical 
effects may outperform classical approaches. In this paper, we propose a new technique 
for locking the frequencies of two distance oscillators, via the process of {\em wavelength 
teleportation}.  

   In this technique, the phase variation of an oscillator is first mapped by Alice (keeper of the 
first clock) to the wave-functions of an array of atoms, via the use of the Bloch-Siegert 
oscillation, which results from an interference between the co- and counter-rotating parts of a 
two-level excitation  \cite{Shahriar02,Cardoso03}. The maximum number of atoms 
needed to encode the phase variation  can be  very small, and is given by the 
Nyquist sampling criterion. Distant entanglement, produced using  an asynchronous 
technique \cite{Lloyd01}, is used to teleport the quantum state of each of 
these atoms to a  matching atom with Bob (keeper of the second clock).  Bob can thus recreate the 
exact phase  variation of Alice's operator locally, and compare with the same for his oscillator. 
We  discuss  the potential constraints and advantages of this approach after presenting the scheme 
in detail.    

  Consider first a situation where Alice and Bob each has an atom that has two degenerate 
ground states ($|0\rangle$ and $|1 \rangle $), each of which is coupled to a higher energy state $(|2 \rangle)$, 
as shown in figure 1. We assume the $0-2$ and $1-2$ transitions are magnetic dipolar, 
and orthogonal to each other, with a transition frequency $\omega$. For example, 
in the case of  $^{87}Rb$, $|0 \rangle$ and $|1 \rangle$ correspond to  $5^{2}P_{1/2}:|F=1,mF=-1 \rangle$ and 
$5^{2}P_{1/2}:|F=1,mF=1 \rangle $ magnetic sublevels, respectively, and $|2 \rangle$ 
corresponds to $5^{2}P_{1/2}:|F=2,mF=0 \rangle $ magnetic sublevel 
\cite{Cardoso03}. Left and right circularly  polarized magnetic fields, 
perpendicular to the quantization axis, are used to excite the 0-2 and 1-2 
transitions, respectively. We take to be the same as the clock frequency $\omega_c$.
\vspace{-.2cm}
\begin{figure}
\epsfxsize=7cm
\epsfysize=5cm
\vspace{-.2cm}
\centerline{\epsfbox{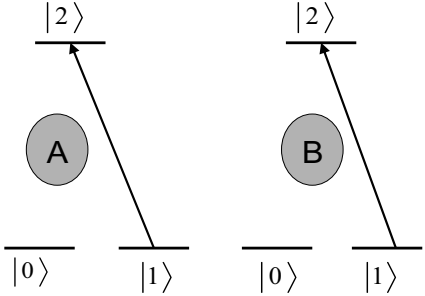}}
\caption{ Schematic illustration of the basic protocol for phase locking two remote clocks, one 
with Alice (A), and the other with Bob (B), without transmitting a clock signal directly. The model 
energy levels can be realized, for example, using the metastable hyperfine Zeeman sublevels of  
$^{87}$Rb 
atoms, as detailed in the text.}
\label{fig_1}
\end{figure}  
  We assume that Alice and Bob's fields at $\omega$  have the form $B_{A}=B_{A0}\cos(\omega t 
+\phi )$ and $B_{B}=B_{B0}\cos(\omega t+\chi )$, respectively. The origin of the time 
variable, t, is therefore arbitrary, and does not affect the   phase 
difference, $\Omega \equiv (\phi -\chi )$. The clocks are assumed to be in phase 
if $\Omega =0$, so that if Bob determines that at some instant his magnetic 
field is maximum and positive in some direction $r_B$, then Alice 
will also find her magnetic field to be maximum and positive in some direction $r_A$ at the same 
instant.  As long as Alice and Bob agree on this definition of phase-locking, and use the 
same definitions all the time, then $r_B$ and $r_A$ do not have to be the same.  
During the magnetic resonance excitations, the value of any dc magnetic field will be assumed 
to be vanishing.  Symmetry then dictates that any physical observable will be independent 
of the choice of the quantization axis, as long as it is perpendicular 
to $r_A$ for Alice, and perpendicular to $r_B$ for Bob.  In order to describe our protocol, we now 
summarize briefly the theory behind the  Bloch-Siegert oscillation that occurs when a 
two-level interaction is considered without the rotating wave approximation (RWA)
\cite{Corney77,Allen75,Bloch40,Shirley65,Stenholm73} and is presented in greater detail 
in ref. \cite{Shahriar02}. We also describe the condition for the time  reversal of an 
arbitrary evolution under this condition, another necessary element of our protocol.  

 We consider an ideal two-level system where a ground state  $|0 \rangle $  is coupled to a higher 
energy state $|1 \rangle $.  We assume that the $1-2$  transition is magnetic dipolar, with a 
transition frequency  $\omega$, and the magnetic field is of the form 
$B=B_{0} Cos(\omega t+\phi)$. In the dipole approximation, the Hamiltonian can 
be written as:
\begin{equation} 
\hat{H} = \epsilon ( \sigma_0 -\sigma_z )/2 + g(t) \sigma_x ,
\label{hmt_original}
\end{equation}
where $g(t) = -g_0\left[\exp (i\omega t + i\phi)+c.c. \right] /2$, 
$\sigma_i$ are the Pauli matrices, and  $\epsilon=\omega$ 
corresponds to resonant excitation. The state vector is written as:
\begin{equation} 
|\xi(t)\rangle = \left( \begin{array}
{c} C_0(t)  \\  C_1(t)
\end{array} \right). 
\label{ket_c_original}
\end{equation}

We  perform a rotating wave transformation by operating on  $|\xi(t)\rangle$
with the unitary operator                                                     
$\hat{Q} = (\sigma_0 + \sigma_z )/2 + 
 \exp (i\omega t + i\phi)(\sigma_0 - \sigma_z)/2.$ 
The Schr\"{o}dinger equation then  takes the form (setting $\hbar=1$):
$\dot{ |\tilde{\xi}\rangle } = -i H(t) |\tilde{\xi}(t) \rangle $ 
where the effective Hamiltonian is given by:
\begin{equation} 
\tilde{H} =  \alpha(t)\sigma_{+} + \alpha^{*}(t) \sigma_{-},
\label{hmt_tilde}
\end{equation}
with $\alpha(t) =  - (g_0/2)\left[\exp (-i2\omega t- i2\phi)+1 \right] $, 
and in the rotating frame the state vector is:
\begin{equation} 
|\tilde{\xi}(t) \rangle  \equiv \hat{Q}|\tilde{\xi}(t)\rangle= \left( 
\begin{array}{c} \tilde{C}_0(t)  \\  \tilde{C}_1(t)
\end{array} \right). 
\label{ket_c_tilde}
\end{equation}
%
%

The general solution without RWA to Eq.4 can be written in the form:
\begin{equation} 
|\tilde{\xi}(t)\rangle = \sum_{n=-\infty}^{\infty}
\left( \begin{array}{c} a_n  \\  b_n \end{array} \right) \exp(  n(-i2\omega t-i2\phi)). 
\label{xi_Bloch_expn}
\end{equation}
with the couplings described by
\begin{mathletters}
\begin{eqnarray}
  \dot{a}_n & = & i2n\omega a_n + i g_0(b_n +b_{n-1})/2, \\
  \dot{b}_n & = & i2n\omega b_n + i g_0(a_n +a_{n+1})/2.
\end{eqnarray}
\label{a_n_b_n_eqn}
\end{mathletters}                 
We consider $ g_0\equiv g_0(t) =g_{0M}(1-e^{-t/\tau_{sw} })$ to have a slower 
time-dependence compared to other characteristic timescales such as $1/\omega$  and 
$1/g_{0M}$, where $g_{0M}$ is the peak value of $g_0$ and $\tau_{sw}$ 
is the switching time. Under this condition, one can solve  these equations by employing the 
method of adiabatic elimination, which is valid to first order
in  $\eta=(g_0/4\omega)$.  As derived in refs. \cite{Shahriar02} and 
\cite{Cardoso03}, the solutions are:
\begin{mathletters}   
\begin{eqnarray}
C_0(t) &=& \cos(g_0'(t)t/2) - 2\eta\Sigma \sin(g_0'(t)t/2), \\
C_1(t) &=& i e^{-i(\omega t + \phi)} [ \sin(g_0'(t)t/2) +  \nonumber \\ 
       &+&2\eta\Sigma^* \cos(g_0'(t)t/2) ],
\end{eqnarray}
\label{c_total_eqn}
\end{mathletters}
\noindent where $\Sigma\equiv(i/2)\exp(-i(2\omega t+ 2\phi))$ and 
$g_0'(t)=1/t \int_0^t g_{0}(t)dt = g_0 \left[1-(t/\tau_{sw})^{-1}\exp(-t/\tau_{sw})\right].$
To lowest order in $\eta$, this solution is normalized at all times. Note that if Alice were  to carry 
this excitation on an ensemble of atoms using $ \pi/2$  pulse and measure the population of the 
state  $|1\rangle$  immediately ( at $t=\tau $, the moment when $\pi/2$ excitation ends), 
the result would be a  signal given by, $\frac{1}{2}\left[1+2\eta \sin(2\omega\tau + 2\phi)\right] $
which  contains information of both the amplitude and the phase of her  field.

  Next, we consider the issue of exact time reversal of such an excitation.  
The Schr\"{o}dinger eqn. (4) has the formal solution:
\begin{equation}
|\tilde{\xi}(t_2) \rangle  =exp(-i\int^{t_2}_{t_1} \tilde{H}(t')dt' |\tilde{\xi} (t_1) \rangle .
\end{equation}

  If the RWA is made, then $\tilde{H}$ is time independent. In that case, if one starts an
evolution at $t_1$, proceed for {\em any} duration T, then reverses the sign of
$\tilde{H}$  by shifting the phase of the magnetic field by $\pi$, and continues with the 
evolution for another duration T, then the system returns back to the starting state.  
Here, however, RWA is not made, so that $\tilde{H}$ depends on time.  
Therefore, the exact reversal can be achieved in this manner only  if $T=m\pi/\omega$  
for any integer value of m  \cite{Jozsa00,Shahriar02,Cardoso03}.

  Returning to the task at hand, our protocol starts by using a scheme, 
developed earlier by us \cite{Lloyd01} 
to produce a degenerate entanglement of the form: 
$|\psi \rangle  = ( |0 \rangle_A |1\rangle_B - |1\rangle_A |0 \rangle_B)/ 
\sqrt{2}$.  
Next, Alice attenuates her field so that the counter-rotating 
term in the Hamiltonian can be ignored (this assumption is not essential for our conclusion, 
but merely simplifies the algebra somewhat), and excites a $\pi$-pulse coupling $|1 \rangle_A$ to 
$|2 \rangle _A$, and then stops the excitation.  Similarly, Bob uses a field, attenuated as above, to 
excite a  $\pi$-pulse coupling $|1 \rangle_B$ to $|2 \rangle_B$, and then stops the excitation.  Using digital 
communications over a  classical channel, Alice and Bob wait until they both know that these 
excitations have been completed. The resulting state is then given by :
\begin{eqnarray}
|\psi(t) \rangle &=&[|0 \rangle _A |2 \rangle_B e^{(-i\omega t-i\chi )} \nonumber\\
  && - |2 \rangle _A |0 \rangle _B e^{(-i\omega t-i\phi ) } ]/\sqrt{2}.
\end{eqnarray}  

The next step is for Alice to make a measurement along the $|0\rangle_A \rightarrow 
|2\rangle _A$ transition. 
For this process, she chooses a much larger value of $g_{0M}$, so that the RWA can not be made. 
The state she wants to measure is  the one that would result if one were to start from state 
$|0 \rangle _A$, and evolve the system for a $\pi /2$ pulse using this stronger $g_{0M}$
\begin{eqnarray}
|+ \rangle_A \equiv &&  (1/\sqrt{2})[(1-2\sigma \Sigma)|0 \rangle_A  \nonumber \\ 
     && +  i e^{-i(\omega t + \phi)}(1+2\sigma \Sigma^*)|2 \rangle_A ],
\end{eqnarray}
where we have made use of eqn.9.  The state orthogonal to $|+\rangle _A$ results from 
a $3\pi /2$ pulse:
\begin{eqnarray}
|- \rangle_A \equiv && (1/\sqrt{2}) [(1+2\sigma \Sigma)|0 \rangle_A \nonumber \\
           && - i e^{-i(\omega t -\phi)}(1+2\sigma \Sigma^*)|2 \rangle _A ].
\end{eqnarray}
%

To first order in $\eta$, these two states are each normalized, 
and orthogonal to each other.  As such, one can re-express the state of the two atoms 
in eqn. 9  as:
\begin{equation}
 |\psi(t) \rangle =(1/\sqrt{2}) (|+ \rangle_A |->_B - |-\rangle _A |+ \rangle_B).  
\end{equation}
where we have defined:
\begin{mathletters}   
\begin{eqnarray}
|+ \rangle _B \equiv && (1/\sqrt{2}) [(1-2\sigma \Sigma)|0 \rangle _B \nonumber \\
      && + i e^{-i(\omega t + \phi)}(1+2\sigma \Sigma^*)|2 \rangle _B ], \\
|- \rangle_B \equiv && (1/\sqrt{2})[(1+2\sigma \Sigma)|0 \rangle_B \nonumber \\
      && - i e^{-i(\omega t -\phi)}(1+2\sigma  \Sigma^*)|2 \rangle_B ].                
\end{eqnarray}
\end{mathletters}   

She can measure the state $|+ \rangle_A$ by taking the following steps: 
(i) Shift the phase of the B-field by $\pi$, (ii) Fine tune the value of $g_0$ 
so that $g'_0(t)=\omega /2m$, for an integer value of m, (iii) apply the field for a 
duration of $T= \omega/2g'_0(T)$, and (iv) detect state $|0\rangle_A$.  
Note that the constraint on $g_0$ ensures that $T=m\pi/\omega$, which is necessary for time 
reversal to work in the absence of the RWA. Once Alice performs this measurement, 
the state for Bob collapses to $|- \rangle_B$, given in Eq. (14).  Note that if $\eta$  is 
neglected, then the measurement produces a $|- \rangle_B$ that contains no 
information about the phase of Alice's clock, which is analogous to the 
Jozsa protocol \cite{Jozsa00}.
 
 In the present case, $|- \rangle_B$  does contain information about the 
amplitude and the phase 
of Alice's clock signal. In order to decipher this, Bob measures his state $|1 \rangle_B$.  
The probability of success is:
\begin{equation}
P_{\phi}\equiv |_{B}\langle 1|- \rangle _{B}|^2 =(1/2) \left[ 1+ 2\eta \sin(2\phi) \right],
\end{equation}
where we have kept terms only to the lowest order in $\eta$. 
Of course, the value of $\phi$ (mod $2\pi$ ), the phase difference, can not be determined 
from knowing $\sin(2\phi )$ alone.  However, this whole process can be repeated after, 
for example, Alice shifts the phase of her B-field by $\pi/2$, so that Bob can determine the 
value of $\cos(2\phi )$. It is then possible to determine the value of $\phi$ (mod $2\phi$ ) 
unambiguously.
\vspace{-.3cm}
\begin{figure}
\epsfxsize=8cm
\epsfysize=6.3cm
\centerline{\epsfbox{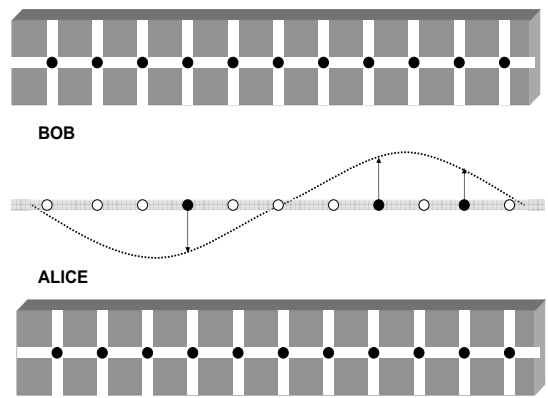}}
\vspace{-.2cm}
\caption{ The top (bottom) array shows the atoms co-located with and excited by Bob (Alice). 
The degree of correlation observed after executing the frequency-locking protocol 
displays a spatial variation only if the frequencies of Bob's and Alice's oscillators are 
different, as shown in the middle.  Elimination  of such a variation leads to frequency locking.}
\label{fig_3}
\end{figure}  

 The overall process can be carried out in one of two ways.  First, consider the situation 
where Alice and Bob starts with X pairs of atoms, and entangle each pair in the form of eqn 13.  
Then, over a digital communication channel, Alice sends Bob a list of the M atoms she found in state 
$|0\rangle _A$ after performing her measurement process described above.  Bob performs his measurement 
only on this subset of atoms.  Suppose he finds L number of atoms in state $|0 \rangle_B$.  Then
\begin{equation}
\zeta \equiv (L/M -1/2) \rightarrow \sin(2\phi),
\end{equation}
for a large M.  Thus, the value of $\zeta$ determined asymptotically for a large number of 
entangled  pairs will reveal the value of $\sin(2\phi )$. Alternatively, if only a single pair of 
atoms is available, then the same result can be obtained by repeating the whole process X times, 
assuming that $\phi$  remains unchanged during the time needed for the process.

	Note that what is determined by Bob is $\phi$, not  $\Omega$.  
Thus, it is not possible to measure the absolute phase difference in this manner.
However, one could use this approach of phase teleportation in order to 
achieve frequency locking of two remote oscillators. This is illustrated in 
figure 2. Briefly, assume that Bob has an array of N atoms.  Assume further that Alice also has an identical 
array of atoms.  For our protocol, the physical separations between the neighboring atoms do not have 
to match.  In principle, one can create such an identical pair of arrays by embedding N rows of atoms 
(or quantum dots) in a substrate patterned lithographically, with two atoms in each row, and then 
splitting it in two halves.  To start with, the corresponding atoms in each array are entangled with 
each other using the asynchronous approach of ref. 10. Here, we assume that the two clocks may differ 
in frequency.  The frequency-locking algorithm then proceeds as follows.  Alice and Bob both apply 
their fields parallel to their arrays of atoms, so that the phase variation is $2\pi$  over their 
respective wavelengths.  After Alice makes her measurements of the state $|+\rangle_A$, using the same set of 
steps as described above, she informs Bob, over a classical communication channel, the indices of her 
atoms that were found in this state.   Bob now measures the state $|-\rangle_B$ for this subgroup of atoms 
only, using an analogous set of time-reversed excitation steps which ends in observing his atom in 
state $|2\rangle_B$. For a given atom in this subgroup, the phase of his field at that location at the 
time Bob starts the measurement affects the probability of success in finding the 
atom in state $|2\rangle_B$ at the end of the measurement process.  
This phase is varied as Bob repeats the measurement different 
measurement-starting-times (modulo $2\pi /\omega_B$, where  B is the frequency of Bob's clock).   
It is easy to show that there exists a choice of this phase for which the probability 
of success is 100$\%$.  However, the success probability for atoms (in the post-selection 
subgroup) would vary with location if the frequencies of Bob's and Alice's clocks are not the same. 
This effect can be used by Bob to adjust his clock frequency, thereby achieving 
frequency locking.  The Nyquist sampling criterion dictates that the number of atoms in this 
subgroup can be as low as only two, so that N can be quite small, thus making this protocol 
potentially practicable. 

 To summarize, previously we have shown how the phase of an 
electromagnetic field can be determined by measuring the population of either of the 
two states of a two-level atomic system excited by this field, via the so-called Bloch-Siegert 
oscillation. 
Here, we show how a degenerate entanglement, created without 
transmitting any timing signal, can be used to teleport this phase information.
This in turn makes it possible to achieve wavelength teleportation, one possible application of 
which is frequency-locking of remote oscillators.

This work was supported by DARPA grant $\#$ F30602-01-2-0546 under the QUIST program, ARO grant 
$\#$ DAAD19-001-0177 under the MURI program, and NRO grant $\#$ NRO-000-00-C-0158.

\end{document}